%% file: paper.tex
\documentclass[sigconf]{acmart}
\usepackage{fancyvrb} % For inline verbatim text
\usepackage{color}
\usepackage{pifont} 
\usepackage{enumitem}
\setlist[itemize]{leftmargin=11pt}

\AtBeginDocument{%
  \providecommand\BibTeX{{%
    \normalfont B\kern-0.5em{\scshape i\kern-0.25em b}\kern-0.8em\TeX}}}

\input{paper-macros}

\input{paper-packages}

\definecolor{rulecolor}{RGB}{255,0,0}
\definecolor{commentcolor}{RGB}{128,128,128} % Grey for comments
\definecolor{placeholdercolor}{RGB}{0,128, 0} % Teal for placeholders

\definecolor{codebg}{rgb}{0.95,0.95,0.95}
\definecolor{keyword}{rgb}{0.0,0.0,0.5}
\definecolor{string}{rgb}{0.5,0.0,0.0}
\definecolor{comment}{rgb}{0.25,0.5,0.25}
\definecolor{function}{rgb}{0.0,0.0,0.6}
\definecolor{number}{rgb}{0.0,0.0,0.6}

\definecolor{mygreen}{rgb}{0,0.6,0}
\definecolor{mygray}{rgb}{0.5,0.5,0.5}
\definecolor{mymauve}{rgb}{0.58,0,0.82}

\lstdefinelanguage{Rust}{
  keywords={fn, let, mut, pub, const, as, use, self, impl, trait, match, enum, struct, ref, if, else, loop, while, for, in, return, break, continue},
  keywordstyle=\color{blue}\bfseries,
  ndkeywords={i32, u32, usize, Result, Option, Some, None, Ok, Err, String, Vec},
  ndkeywordstyle=\color{mymauve}\bfseries,
  identifierstyle=\color{black},
  sensitive=true,
  comment=[l]{//},
  morecomment=[s]{/*}{*/},
  commentstyle=\color{mygreen}\ttfamily,
  stringstyle=\color{mymauve}\ttfamily,
  morestring=[b]',
  morestring=[b]",
}

\lstdefinelanguage{C}{
  language=C,
  morekeywords={uint8_t, uint16_t, uint32_t, int8_t, int16_t, int32_t, size_t},
  keywordstyle=\color{blue}\bfseries,
  ndkeywords={NULL},
  ndkeywordstyle=\color{mymauve}\bfseries,
  identifierstyle=\color{black},
  sensitive=true,
  comment=[l]{//},
  morecomment=[s]{/*}{*/},
  commentstyle=\color{mygreen}\ttfamily,
  stringstyle=\color{mymauve}\ttfamily,
  morestring=[b]',
  morestring=[b]",
}

\microtypecontext{spacing=nonfrench}

\hypersetup{
  bookmarks=false, % don't show bookmarks tab on startup
  colorlinks=true,%
  citecolor=black,%
  filecolor=black,%
  linkcolor=black,%
  urlcolor=black,%
  pagebackref=true,%
}

\copyrightyear{2026}
\acmYear{2026}
\setcopyright{cc}
\setcctype{by}
\acmConference[ReCode '26]{1st Workshop on Code Translation, Transformation, and Modernization }{April 12--18, 2026}{Rio de Janeiro, Brazil}
\acmBooktitle{1st Workshop on Code Translation, Transformation, and Modernization (ReCode '26), April 12--18, 2026, Rio de Janeiro, Brazil}
\acmPrice{}
\acmDOI{10.1145/3786180.3788317}
\acmISBN{979-8-4007-2411-4/2026/04}

\begin{document}

%\title{From C to Rust: Evaluating LLMs in Code Translation and Vulnerability Mitigation}
\title{SafeTrans: LLM-assisted Transpilation from C to Rust}
% Authors
\input{0a-authors.tex}
% \author{Anonymous Authors}

% Abstract and CCS concepts

\input{0d-ccs}

\input{0c-abstract}
% Start the main content
\maketitle
% \pagestyle{plain}
% Paper sections

\input{introduction.tex}

\input{related-work}
\input{methodology.tex}

\input{experimental-setup.tex}
\input{results.tex}

\input{vulnerability.tex}

\input{limitations}

\input{conclusion.tex}

\begin{acks}
We thank the anonymous reviewers for their constructive feedback. This
work was supported by the Office of Naval Research (ONR) through award
N00014-24-1-2054, with additional support by Amazon
through the Amazon Research Awards program.
\end{acks}

% Bibliography TODO: Uncomment it when citations are added
\bibliographystyle{ACM-Reference-Format}
\newpage
\bibliography{references}

% \clearpage
\newpage
\appendix
\input{appendix.tex}
\end{document}

%% file: paper-macros.tex
\definecolor{lightgrey}{rgb}{0.8,0.8,0.8}
\definecolor{white}{rgb}{1.0,1.0,1.0}

%% file: paper-packages.tex
\usepackage{makecell}
\usepackage{array}
\usepackage{xspace}
\usepackage{enumitem}
\setlistdepth{9}

\usepackage{subcaption}
\usepackage{listings}

%% file: 0a-authors.tex
% \author{Anonymous\\ Authors} 

% \author{Reinhard Sch\"utte}
%     \affiliation{%
%         \institution{Institute for Computer Science \\ University of Duisburg-Essen}
%         \streetaddress{}
%         \city{Essen}
%         % \state{}
%         \country{Germany}
%         \postcode{45141}
%     }

\author{Muhammad Farrukh}
    \affiliation{%
        \institution{Stony Brook University}
        \city{New York}
        \country{USA}
    }
    \email{mufarrukh@cs.stonybrook.edu}

\author{Baris Coskun}
    \authornote{This work does not relate to Baris Coskun's position at Amazon.}
    \affiliation{%
        \institution{Amazon Web Services}
        \city{New York}
        \country{USA}
    }
        \email{barisco@amazon.com}

\author{Tapti Palit}
    \affiliation{%
        \institution{University of California, Davis}
        \city{Davis}
        \country{USA}
    }
    \email{tpalit@ucdavis.edu}

\author{Michalis Polychronakis}
    \affiliation{%
        \institution{Stony Brook University}
        \city{New York}
        \country{USA}
    }
    \email{mikepo@cs.stonybrook.edu}

\renewcommand{\shortauthors}{Farrukh et al.}

%% file: 0d-ccs.tex
    
\ccsdesc[300]{Security and privacy~Software security engineering}

\keywords{Transpilation, Rust, LLM, Memory Safety, Vulnerability Mitigation}

%% file: 0c-abstract.tex
\begin{abstract}

Rust is a strong contender for a memory-safe alternative to C as a
``systems'' language, but porting the vast amount of existing C code to Rust remains daunting.
In this paper, we evaluate the potential of large language models (LLMs) to
automate the transpilation of C code to idiomatic Rust.
We present \emph{SafeTrans}, a generic framework that leverages LLMs to
i)~transpile C code into Rust, and ii)~iteratively repair compilation and runtime errors.
A key novelty of our approach is a few-shot \emph{guided repair} technique for translation errors,
which provides contextual information and example code snippets for specific error types, guiding the LLM toward the correct solution.
Another novel aspect of our work is the evaluation of the security implications of the transpilation process, showing how some vulnerability classes in C persist in the translated Rust code.
SafeTrans was evaluated with six leading LLMs on 2,653 C programs and two real-world C projects.
Our results show that iterative repair improves the rate
of successful translations from 54\% to 80\% for the best-performing LLM (\texttt{gpt-4o}).

\end{abstract}

%% file: introduction.tex
\section{Introduction}
\label{sec:introduction}

Many of the software systems used by enterprises and individual users rely on
huge legacy C/C++ code bases. Migration to memory-safe languages will be a slow
and tedious task if not (at least partially) automated. Rust is the strongest
contender for a memory-safe ``systems'' language with acceptable runtime
overhead, and major projects 
have started distributing some components written in Rust,
or at least have introduced tooling support for integrating
Rust code in existing C/C++ code bases
(e.g., Firefox, Chrome, Linux, Windows).
Automating the translation of \emph{existing} C code into Rust,
however, is challenging due to substantial syntactic and semantic disparities
between the two languages, particularly concerning memory management and
ownership.

Initial attempts to this problem adopted rule-based translation
approaches~\cite{immunant2022c2rust, crown, hong2023concrat, hong2024tag,
safer-rust}. These methods scale to large programs and yield functionally
equivalent code, but often fail to produce idiomatic and safe Rust, undermining
their security benefits.
% . More importantly, the majority of the translated code is
% wrapped in \texttt{unsafe} blocks (Rust's way of allowing developers to bypass
% its safety guarantees), which defeats the purpose
% %of translation
% from a security
% perspective. To address the limitations of rule-based methods, learning-based
% transpilers have been proposed, which convert code translation into a neural
% machine translation problem~\cite{chen2018tree, lachaux2020unsupervised,
% roziere2021leveraging, szafraniec2022code}. These techniques offer improvements
% over traditional program analysis methods, but bring new challenges, such as the
% high amount of resources required for training, and the scarcity of functionally
% equivalent program pairs in source and target languages like C and Rust,
% respectively.

%In a relatively short span, the fields of artificial intelligence and natural language
%processing have achieved impressive advances in generative AI, with a multitude of LLMs trained
%on various sources of data and for a wide range of applications, including images, text, and code.
%In the field of computer programming, in particular, LLMs have shown great promise for code
%auto-completion, synthesizing code from natural language descriptions, summarizing and explaining
%existing code, and various other code-related tasks. By training models on immense code bases,
%such as source code files from public GitHub repositories, these code-specific LLMs can learn rich
%contextual representations that can be applied to various code-related tasks.
With the advent of large language models (LLMs), recent studies have focused on
exploring their potential for automating code translation. LLMs can generate
more idiomatic code than previous methods, but they come with their own set of
challenges. Pan et al.~\cite{pan2024lost} performed one of the first studies to
understand the limitations of LLMs for code translation, and present a
taxonomy of bugs introduced during the translation process. Similarly, Ou et
al.~\cite{ou2024repository} developed a repository-level benchmark
%(RustRepoTrans)
for code translation evaluation, and developed a taxonomy of
translation errors. LLM-transpiled code requires comprehensive methods to
verify its correctness, an issue that recent studies attempt to address using various
techniques~\cite{yang2024unitrans, eniser2024flourine, nitin2024spectra, yang2024vert,bai2025rustassure}.
% automated test case generation~\cite{yang2024unitrans},
% fuzzing~\cite{eniser2024flourine}, multimodal
% specification~\cite{nitin2024spectra}, formal
% verification~\cite{yang2024vert}.
Aside from correctness, LLMs also struggle with larger
applications. Modern LLMs support large context windows, but recent studies
show that they often cannot attend to all parts of long inputs uniformly,
 impacting their effectiveness~\cite{10.1162/tacl_a_00638,li2024long}. Several studies have tackled this issue
by dividing larger applications into smaller translation
tasks~\cite{nitin2025c2saferrust, shetty2024syzygy, shiraishi2024context,
zhang2024scalable}. 
% Although these studies explain the causes of translation
% errors, it remains unclear how to leverage this knowledge to improve
% transpilation accuracy.
Although these studies identify the causes of translation errors, how to use this knowledge to improve transpilation accuracy remains unclear.

Unlike the above recent works in automated transpilation using LLMs,
which mostly focus on
handling the translation of larger applications and developing additional
approaches to verify the correctness of the translated programs, our research
addresses several fundamental questions specific to C-to-Rust translation.
First, we identify the Rust features that LLMs struggle to handle, which
lead to the most frequent translation errors. We extend our analysis beyond building a
simple taxonomy, by incorporating these findings to improve translation
accuracy in the form of
%concrete examples and
contextual information that aids the repair process.

Second, we evaluate the effectiveness of using this additional contextual
information as part of a \emph{guided} repair process to
fix faulty translations.
This is achieved by
analyzing the most frequent translation errors and assembling targeted
instructions along with example code snippets to guide the LLMs towards the
correct solution.
Finally, we examine the effect of vulnerabilities present in C on the corresponding Rust translations.
This targeted approach distinguishes our work from previous transpilation
studies by specifically examining the language-level semantic challenges in
translating from a memory-unsafe language to one with strict safety guarantees
enforced at compilation time.

To this end,
% and aiding the migration to memory-safe languages, in this paper
 we developed \emph{SafeTrans},
a framework for evaluating LLMs in their C-to-Rust code translation capabilities.
% %an LLM-based framework for automating the rewriting of existing C/C++ code into Rust.
% SafeTrans uses LLMs to i)~transpile C code into Rust
% and ii)~iteratively fix any compilation and runtime errors in the resulting code.
% Successfully transpiled programs are then verified against the original C program's unit tests
% to ensure their functional correctness.
We used SafeTrans to perform a total of 15,918 translations across
2,653 C programs and six LLMs.
The combination of basic repair with few-shot guided repair
achieves compilation repair success rates of up to 93.5\% for \texttt{gpt-4o} and 89.8\%
for \texttt{DeepSeek-V3}. Guided repair effectively resolves
challenging Rust compilation errors, such as trait implementation
failures, with an average resolution rate of 58.7\%, and
borrow-checker violations with an average rate of 74.2\% across all LLMs. Overall, our
repair techniques collectively achieve 
substantial improvements in computational accuracy (CA), increasing the
overall translation success rate from 54\% to
80\% for \texttt{gpt-4o}. Even smaller models, such as \texttt{Qwen2.5-Coder} and \texttt{DeepSeek-Coder}, nearly
double their CA, highlighting the broad applicability and effectiveness of our
approach.

In summary, we make the following main contributions:
\begin{itemize} 
    \item We present the design and implementation of SafeTrans, an end-to-end
    framework for comprehensively evaluating the C-to-Rust code transpilation capabilities of LLMs.

    \item We demonstrate that just providing compiler error messages and
    feedback is
    insufficient for repairing many types of faulty translations, and introduce a novel
    few-shot guided repair approach to improve the repair rate.  
    \item We identify 10,375 vulnerabilities (e.g., out of bounds access,
    null-pointer dereference) in the 2,653 source C programs
    used in our evaluation,
    and demonstrate their behavior in the translated
    Rust programs.
\end{itemize}
Our prototype implementation and data set are publicly available through
\url{https://github.com/FarrukhCyber/SafeTrans}.

%% file: related-work.tex
\section{Background and Related Work}
\label{sec:related_work}

% Source-to-source code translation is a decades-old problem in the programming
% languages and software engineering communities~\cite{ada-pascal}, driven by the
% need to modernize applications, migrate legacy systems, and leverage the
% benefits of newer languages.
% Translation is achieved using a
% source-to-source compiler, also known as a 
% \emph{transcompiler} or \emph{transpiler}, i.e., a program that
% converts between programming languages
% that operate at a similar level of abstraction~\cite{roziere2020unsupervised}.
% Translating C programs to Rust has received significant attention due
% to Rust's memory safety and performance characteristics,
% which position it as a safer alternative to C.

% \subsection{Rule-based Code Translation}
\textbf{Rule-based C-to-Rust Translation:}
Conventional solutions to source-to-source translation
have predominantly relied on rule-based methodologies.
% These approaches rely on static analysis to generate the abstract syntax tree
% and control flow graph of the code.
% Carefully crafted rules are then derived mostly manually to transcribe
% the source code into the target language.
% The development of such rule-based techniques is a tedious process that
% involves substantial manual human effort.
% Existing solutions for automating the conversion
% of C code to Rust do not apply
% any of its borrowing and ownership features, resulting in overuse of
% Rust's \texttt{unsafe} keyword in the translated code.
%---thus offering no meaningful security improvement~\cite{c2rust-study}.
The most prominent tool in this category is
C2Rust~\cite{immunant2022c2rust}, which translates C programs to
Rust using both predefined and custom rules. Despite its scalability,
C2Rust produces non-idiomatic code with excessive use of
\texttt{unsafe} blocks. 

A recent study of C2Rust
by Emre et al.~\cite{safer-rust} investigates the underlying causes of
unsafety. The authors propose a technique that relies on 
feedback from the \texttt{rustc} compiler
to refactor a certain type of raw pointers into Rust references.
Inspired by C2Rust, several studies have attempted to address its limitations.
CROWN~\cite{crown} improves upon C2Rust's output by converting raw pointers to
references, but it is limited to mutable and non-array raw pointers. Similarly,
other tools~\cite{hong2023concrat,hong2024don,hongc2rust} focus on specific translation
challenges, such as converting lock APIs and certain data types.

\begin{figure*}[t]
    \centering
    \includegraphics[width=0.88\textwidth]{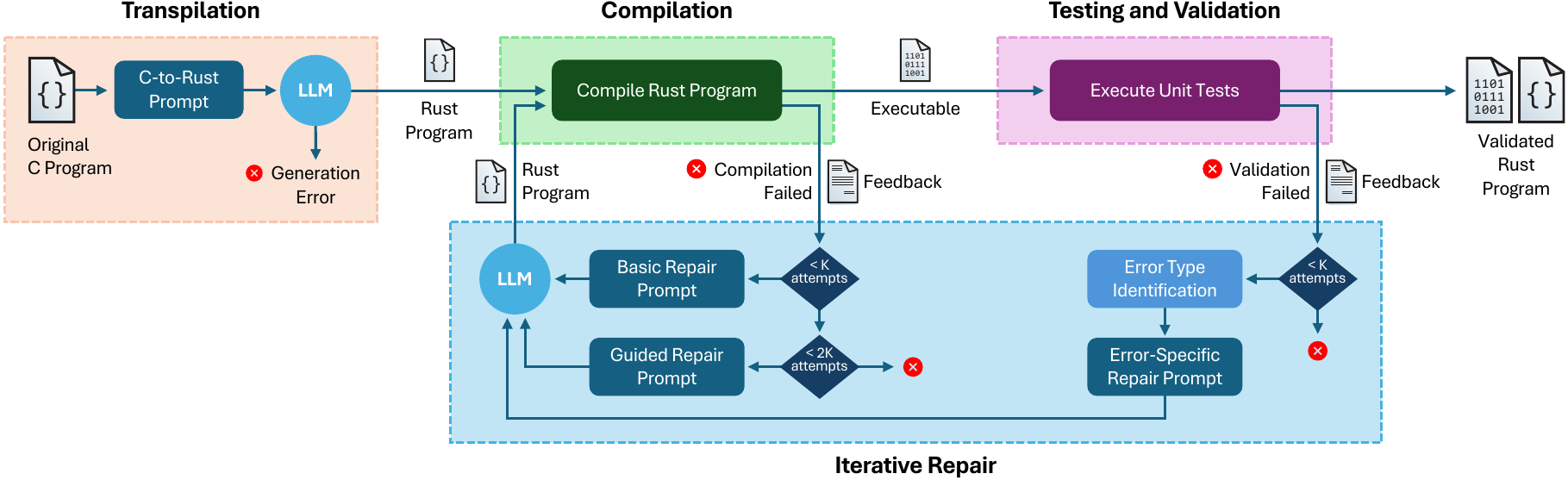}
    \caption{High-level architecture of SafeTrans'
    transpilation, compilation, repair, and validation pipeline.}
    \Description{A flowchart showing the SafeTrans pipeline with four main stages: transpilation of C code to Rust using an LLM, compilation of the generated Rust code, repair of compilation and runtime errors through iterative LLM feedback, and validation using test cases.}
    \label{fig:pipeline}
\end{figure*}

% \subsection{LLM-based Code Translation}
% In a relatively short span, the fields of artificial intelligence and natural
% language processing have achieved impressive advances in generative AI, with a
% multitude of LLMs trained on various sources of data and for a wide range of
% applications
% \cite{HUSEIN2025103917, xu2024large, bozkir2024embedding, kim2025explainablexrunderstandinguser,
% sapkota2025image, muzammil2025babylon, jelodar2025large}. In the field of
% computer programming, in particular, LLMs have shown great promise for code
% auto-completion, synthesizing code from natural language descriptions,
% summarizing and explaining existing code, and various other code-related tasks.
% By training models on immense code bases, such as source code files from public
% GitHub repositories, these code-specific LLMs can learn rich contextual
% representations that can be applied to various code-related tasks.
%With the rise of LLMs, they have become
%increasingly popular in software engineering due to their strong understanding
%of programming syntax across multiple languages.
% This makes them an excellent
% tool for automated code translation~\cite{roziere2020unsupervised}.
\textbf{LLM-based C-to-Rust Translation:}
In recent years, LLMs have been used for a wide range of applications
\cite{HUSEIN2025103917, xu2024large, bozkir2024embedding,
kim2025explainablexrunderstandinguser, sapkota2025image, muzammil2025babylon,
jelodar2025large} including code generation. Pan et al.~\cite{pan2024lost}
conducted a comprehensive analysis of LLMs' translation capabilities across
multiple language pairs, including C to Rust. Their study also provides a
detailed taxonomy of LLM-based translation bugs but lacks in-depth insights into
issues specific to C-to-Rust translation. For C-to-Rust translation, several
studies enhance LLM-based approaches with program analysis, formal verification,
and fuzzing to improve translation accuracy. VERT~\cite{yang2024vert} uses
WebAssembly to generate oracle Rust programs for formal equivalence testing.
FLOURINE~\cite{eniser2024flourine} verifies translated code via fuzzing, and
SPECTRA~\cite{nitin2024spectra} enhances translation using multimodal
specifications. These tools, however, are limited to small programs (<600 LoC).
Some recent works~\cite{nitin2025c2saferrust, shetty2024syzygy, shiraishi2024context, bai2025rustassure, evoc2rust, alphatrans} focus on translating large-scale C programs to Rust by decomposing the source code for incremental translation, while others employ LLM-powered agents for transpilation and testing~\cite{sim2025largelanguagemodelpoweredagent, li2025adversarial, matchfixagent}
% Some recent works~\cite{nitin2025c2saferrust,
% shetty2024syzygy,shiraishi2024context,bai2025rustassure, evoc2rust, alphatrans} focus on translating large-scale C
% programs to Rust with the common approach of first decomposing the source
% program and then performing incremental translation. 
% In parallel, LLM-powered agents are also being used for transpilation and testing~\cite{sim2025largelanguagemodelpoweredagent,li2025adversarial, matchfixagent}. 
% Nitin
% et al.~\cite{nitin2025c2saferrust} propose C2SaferRust, which builds on
% C2Rust transpiler and leverages LLM to reduce unsafe consturcts.
% SYZYGY~\cite{shetty2024syzygy} also presents a method for translating entire
% repositories from C to Rust. Their approach segments the program into smaller
% translation units, uses an LLM to translate each unit, and verifies correctness
% through dynamic analysis and LLM-generated tests.
% Shiraishi et al.~\cite{shiraishi2024context} adopt a similar methodology for handling large C
% programs, but focus solely on generating compilable Rust code without providing a
% means to verify functional correctness. 
Finally, a user study by Li et
al.~\cite{li2024userstudy} demonstrates that human strategies for C-to-Rust
translation differ from those used by automated tools.
% , suggesting that future
% translators could benefit from incorporating human-like decision-making.

% Compared to the above works, which primarily focus on translation accuracy and correctness,
% a key novel aspect of our research is the evaluation of the security implications
% of the transpilation process, i.e., whether potential vulnerabilities in the
% original C code have been properly addressed in the translated Rust code.

%% file: methodology.tex
\section{SafeTrans Overview} 
\label{sec:methodology}

In this section, we present SafeTrans, our framework for automating C-to-Rust
translation, along with the methodology used to evaluate translation fidelity.
The overall translation process comprises four main steps, as illustrated in
Figure~\ref{fig:pipeline}:
transpilation, compilation, repair, and validation.
The original C program is first transpiled into Rust using an 
LLM. The resulting Rust program is then compiled, and if compilation
fails, an iterative repair process is initiated to fix the errors using both
generic and guided prompts.
After successful compilation, the program is executed and validated against
the original C program's test cases. In case of runtime or validation errors,
another iterative repair cycle is initiated.

\subsection{Transpilation from C to Rust}
The first step in the translation pipeline is to generate an initial Rust version of the original C program using an LLM. The input program is included in the prompt, along with instructions for producing safe code and adhering to the required output format.
% The first instruction in the prompt ensures that the LLM produces safe Rust by
% avoiding the generation of \texttt{unsafe} code blocks in the translation.
% The second instruction helps to enforce the inclusion of all necessary dependencies
% in the resulting code. The final two instructions assist in extracting the
% translated code from the LLM's response. This is particularly useful for smaller
% open-source models, which often lack consistency in their output format. To mitigate
% this issue, we explicitly instruct the model to output only code, and wrap it within
% the defined tags.
% Detailed examples of all prompts discussed in this section can be found in Appendix~\ref{sec:templates}.

We use this initial prompt to query the LLM, which can result in two
scenarios.
Ideally, the LLM generates a response that adheres to the
required output format, allowing us to extract the Rust code based on the
predefined tags.
%The extracted Rust code then advances through the pipeline.
In the second case, the LLM fails to generate a valid response, either because the
prompt and the output exceed its context window, or because it produces incoherent
output, without adhering to the required format. In such cases, we
discard the response, a condition we refer to as a \textit{Generation Error}.

% \begin{figure}[t]
%     \begin{tcolorbox}[prompt box]
% Given some code written in the C programming language, translate it into
% equivalent Rust code that solves the exact
% same problem as the original code does. Ensure the following:

% - Produce only safe Rust code.\\
% - The translated Rust code can be compiled and executed with all the necessary
% imports.\\
% - Output only the code without any additional explanation or comments.\\
% - Wrap the code with \Verb_```rust_\\
% C code:\\
% \placeholder{C code}\\
% Rust code:
%     \end{tcolorbox}
%     \caption{Prompt template for C to Rust transpilation.}
%     \label{fig:base_prompt_template}
% \end{figure}

\vspace{-2ex}
\subsection{Compilation and Repair:} 
In the compilation phase,
the transpiled code
%from the previous phase
is compiled using the
\texttt{rustc} compiler. If the compilation is successful, the
resulting binary is provided as input to the runtime testing and validation phase.
Otherwise, SafeTrans attempts to automatically repair the compilation errors
%in the generated Rust code
by initiating an iterative repair phase.

\textbf{Basic Repair:}
In case of a compilation error,
%we receive the erroneous program along with
%\textit{rustc} feedback from the Compile Validation Phase.
the \texttt{rustc}
compiler provides detailed error messages which can be
passed to the LLM along with the transpiled Rust program to provide additional
context about the issue. Recent studies \cite{deligiannis2023fixing} have
demonstrated the effectiveness of this approach.
We adopt an iterative repair approach to resolve compilation errors similar to
Pan et al.~\cite{pan2024lost}. In each iteration, the incorrect Rust translation
and the repair instructions within the prompt are updated based on the
results of the previous iteration, while the base prompt remains unchanged. At
the end of each iteration, the generated Rust program is compiled again,
and if compilation fails, a new repair cycle is initiated.
The phase continues until either the program
is successfully compiled, or a predefined maximum number of iterations is
reached (set to five in our experiments).
% Our repair prompt contains complete contextual information structured into three
% major parts, as illustrated in \figref{fig:repair_prompt}.\farrukh{check it later} The base prompt
% section, which corresponds to the base translation template of
% \figref{fig:base_prompt_template}, helps the LLM
% retain contextual understanding and ensures that the repaired Rust code remains
% equivalent to the core functionality of the original C program. Next, we include
% the incorrect Rust translation generated by the LLM. Finally, we append the
% compilation error messages along with additional instructions about the required
% output format.

% \begin{figure}[t]
%     \begin{tcolorbox}[prompt box]

% \placeholder{Base prompt}
% \promptcomment{// Base transpilation prompt from \figref{fig:base_prompt_template}}\\

% Rust code: \\
% \placeholder{Rust code}
% \promptcomment{// Rust code with compilation errors} \\

% Executing your generated code gives the following errors because it is
% syntactically incorrect:
% \placeholder{error messages} \\
% Please suggest a corrected version of the complete code wrapped in \Verb_```rust_
%     \end{tcolorbox}
%     \caption{Prompt template for repairing compilation errors. }
%     \label{fig:repair_prompt}
% \end{figure}
\textbf{Guided Repair:}
Even after several iterations, some programs may still
fail to compile. To improve the translation success rate,
we introduce a new \emph{guided repair} strategy,
which uses few-shot learning by incorporating error-specific contextual
information in the repair prompt.
% The customized prompt includes 
% specific guidance and concrete code
% examples tailored to the particular compilation errors encountered. %(Appendix~\ref{sec:guidedprompt}).
% During our preliminary experiments, we analyzed the
% frequency and distribution of compilation errors (based on their unique numbers
% as returned by \texttt{rustc}), and identified the most common translation
% errors (discussed in Section~\ref{sec:results}). The error messages
% generated by \texttt{rustc} provide rich information about the root cause of the
% failure and potential solutions.
We selected the top eight most frequent errors,
for which we developed tailored repair instructions, outlining key aspects to
address along with common causes and fixes. 
To maintain conciseness within the LLM's context length
constraints and minimize noise, we only include instructions relevant to the
errors present in the current compilation output.
% when an error is encountered.
The instructions first explain the Rust
property that was violated, leading to that error, followed by
concrete examples of incorrect and corrected code snippets to assist the LLM
in resolving the issue.
% %  (Appendix~\ref{sec:guidedprompt}).
% \figref{fig:guided_repair_inst_example} shows an example of the
% error-specific instructions for error code E0384
% (\emph{``Cannot assign to an immutable variable''}) that are included in the prompt
% when this error is encountered.
% The instructions first explain the Rust
% property that the LLM-generated code violates, leading to error E0384. They then
% provide concrete examples of incorrect and corrected code snippets to help the LLM
% understand and resolve the issue effectively.
\vspace{-2ex}

\subsection{Runtime Testing and Validation} 

Successfully compiled Rust programs proceed to the runtime testing and
validation phase, which executes the program with various test cases and
validates the output against the expected results.
We consider a C program as
successfully translated if the generated Rust program passes all test cases.
Erroneous outcomes of this dynamic analysis phase include
\textit{Runtime Error}, \textit{Infinite Loop}, and \textit{Test Case Error}.
Our unit tests are based on
the test cases available in the CodeNet data set used in our evaluation. If the
translated Rust program fails to run properly or does not pass the test
cases, it undergoes another round of iterative repair, this time with prompts
tailored to the specific type of runtime error encountered.

% First, we identify the type of failure,
% %i.e., \textit{Runtime Error}, \textit{Infinite Loop}, or \textit{Test Case Error}.
% and based on the error type, we construct a dynamic prompt that incorporates the
% corresponding error feedback, 
% % as illustrated in
% % \figref{fig:dynamic_repair_prompt_template}.
% We follow a structure similar to
% the compilation error repair prompts, but this phase handles multiple error
% types at the same time.
% %The new LLM-generated Rust code is sent to Run \& Unit-Test Validation Phase.
% While trying to repair the Rust program, it is possible for the LLM to introduce
% new compilation errors, which will lead to failed validation and the program
% will need to be repaired again. However, only basic repair will be used to
% repair intermediate compilation errors.

In this phase, SafeTrans iteratively repairs the program until it passes all test cases
or exceeds the maximum number of repair attempts (set to five in our tests).
In each iteration, SafeTrans performs both
compilation and runtime test checks, and if the validation fails, it queries the LLM with
an appropriately updated (dynamically generated) prompt.

%% file: experimental-setup.tex
\section{Experimental Setup}
\label{sec:experimentalSetup} 

\textbf{Large Language Model Selection:}
For our empirical study, we select a diverse set of LLMs, ranging from small
open-source models to state-of-the-art (SOTA) LLMs. Among the SOTA LLMs, we
include \texttt{gpt-4o} and \texttt{DeepSeek-V3} (671B). For small open-source LLMs, we
select \texttt{Qwen2.5-Coder} (7B), \texttt{Codestral} (22B), \texttt{DeepSeek-Coder} (16B), and \texttt{Llama3} (70B).
As a rapidly evolving field, LLMs are frequently updated, and new models are
being released regularly. Due to time and cost reasons we could not include
other recently released models, such as Claude and Gemini,
but we tried to select a set of
models that are representative of the spectrum of choices and capabilities
in the current state of the art.
\textbf{Data Set Collection and Pre-Processing:}
% For code translation tasks, prior studies have commonly utilized data sets such
% as CodeNet~\cite{puri2021codenet}, AVATAR~\cite{ahmad2021avatar}, and
% EvalPlus~\cite{liu2023evalplus}. Our empirical study specifically targets the
% translation of C programs to Rust, which necessitates a data set with a
% substantial number of C programs. Having some strong evidence about the validity
% of the translation is also critical, and can be accomplished by respective test
% case inputs and outputs for runtime verification of functional correctness.
% Based on these requirements, CodeNet is the most suitable choice. 
For code translation tasks, we use the CodeNet data
set~\cite{puri2021codenet}, and two real-world
C libraries, \texttt{url\_parser}~\cite{url-parser} and \texttt{avl\_tree}~\cite{avl-tree},
which have been used in prior LLM-based translation studies.
These two programs include comprehensive end-to-end test cases and represent
challenging translation tasks.
The CodeNet data set~\cite{puri2021codenet} comprises 4,053 competitive
programming problems written in over 50 programming languages.
%with approximately 13 million code submissions.
%Each problem includes multiple
%solutions across different languages.
Since our focus is on C, we filter the
data set to retain only those problems that have an adequate number of
solutions written in C.
% resulting in reduced set of 2,653 unique problems. We
% We further eliminate duplicates and randomly sample one solution per problem.
% Additionally, we ensure that all selected problems have a comprehensive set
% of verified test cases.
% %
% Upon closer inspection, we observe that some solutions contain additional
% utility functions that are not called anywhere in the program. These ``dead''
% functions may affect the accuracy of the translation, and therefore we remove them
% using
% %the static analysis tool
% \texttt{tree-sitter}~\cite{tree-sitter}.
% After the above filtering and preprocessing steps,
% our final evaluation data set contains 2,653 C programs.

After some preprocessing steps, our final evaluation data set contains 2,653 C
programs. The test cases in our dataset achieve an average line coverage of
approximately 87\%.

%% file: results.tex
\section{Results}
\label{sec:results}

%%%%%%%%%%%%%%%%%%%%%%%%%%%%%%%%%%%%%%%%%%%%%%%%%%%%%%%%
%%%%%%%%%       TABLES %%%%%%%%%        
%%%%%%%%%%%%%%%%%%%%%%%%%%%%%%%%%%%%%%%%%%%%%%%%%%%%%%%% 

%#CFT: Number of Compilation Failed Translations
%#RT: Number of Compilation Failed Translations

%%%%%%%%%%%%%%%%%%%%%%%%%%%%%%%%%%%%%%%%%%%%%%%%%%%%%%%%

We present the results of our experimental evaluation, focusing on the following research questions:
\textbf{RQ1:} How base LLMs perform on C-to-Rust translation task?
\textbf{RQ2:} What compilation errors occur and how effective is iterative repair?
\textbf{RQ3:} How effective is guided repair for persistent errors?
\textbf{RQ4:} What is the overall improvement in translation success rate?
\textbf{RQ5:} How does SafeTrans scale on complex real-world programs?
   %  and present the trade off of each technique.

%\begin{itemize}
%    \item \textbf{RQ1: Effectiveness of basic LLM-based C to Rust translation.}
%    We evaluate in detail how recent LLMs perform on the task
%    of translating C programs into Rust, and analyze the different failure
%    conditions that are encountered.
%    \item \textbf{RQ2: Analysis of compilation errors.}
%    We explore the different types of compilation
%    errors encountered in the transpiled Rust programs, and evaluate
%    the effectiveness of simple iterative repair.
%    \item \textbf{RQ3: Effectiveness of guided repair.} We investigate how compilation
%    errors evolve and transition during repairing iterations, and evaluate the
%   effectiveness of our guided repair strategy in fixing these errors.
%    \item \textbf{RQ4: Improvement in overall successful translation rate.}
%    We evaluate how our iterative repair strategies help in increasing the
%    number of successfully translated programs.
%    \item \textbf{RQ5: SafeTrans scalability on complex programs.}
%    We compare SafeTrans performance on larger, more complex C applications
%    compared to existing LLM-based transpilers and presents the trade off of each technique.
%\end{itemize}

% \begin{figure}
%     \centering
%     % \includegraphics[scale=0.5 ]{images/methodology.pdf}  % Adjust width as needed
%     \includegraphics[width=0.47\textwidth]{images/base_translation_results.pdf}  % Adjust width as needed
%     \caption{Base Translation Accuracy and Errors Breakdown Across LLMs}
%     \label{fig:base_results}
% \end{figure}

\subsection{RQ1: Basic Translation Success Rate}

To assess the ``out-of-the-box'' performance of LLMs in C to Rust
translation, we adopt \emph{computational accuracy (CA)}, proposed by Rozière et
al.~\cite{roziere2020unsupervised} and Szafraniec et
al.~\cite{szafraniec2022code}, as our primary evaluation metric. CA is defined as
the ratio of successfully translated programs to the total number of translation
samples. 
% We prioritize CA over static evaluation metrics such as exact match,
% syntax match, and dataflow match~\cite{ren2020codebleu}, because CA evaluates the
% functional equivalence of translated programs by executing them with similar
% inputs. LLMs can achieve high scores on static metrics while demonstrating poor
% performance in computational accuracy.
% thereby revealing the limitations of such
%metrics in programming language translation tasks.

\begin{figure}[t]
    \centering
    \includegraphics[width=0.96\columnwidth]{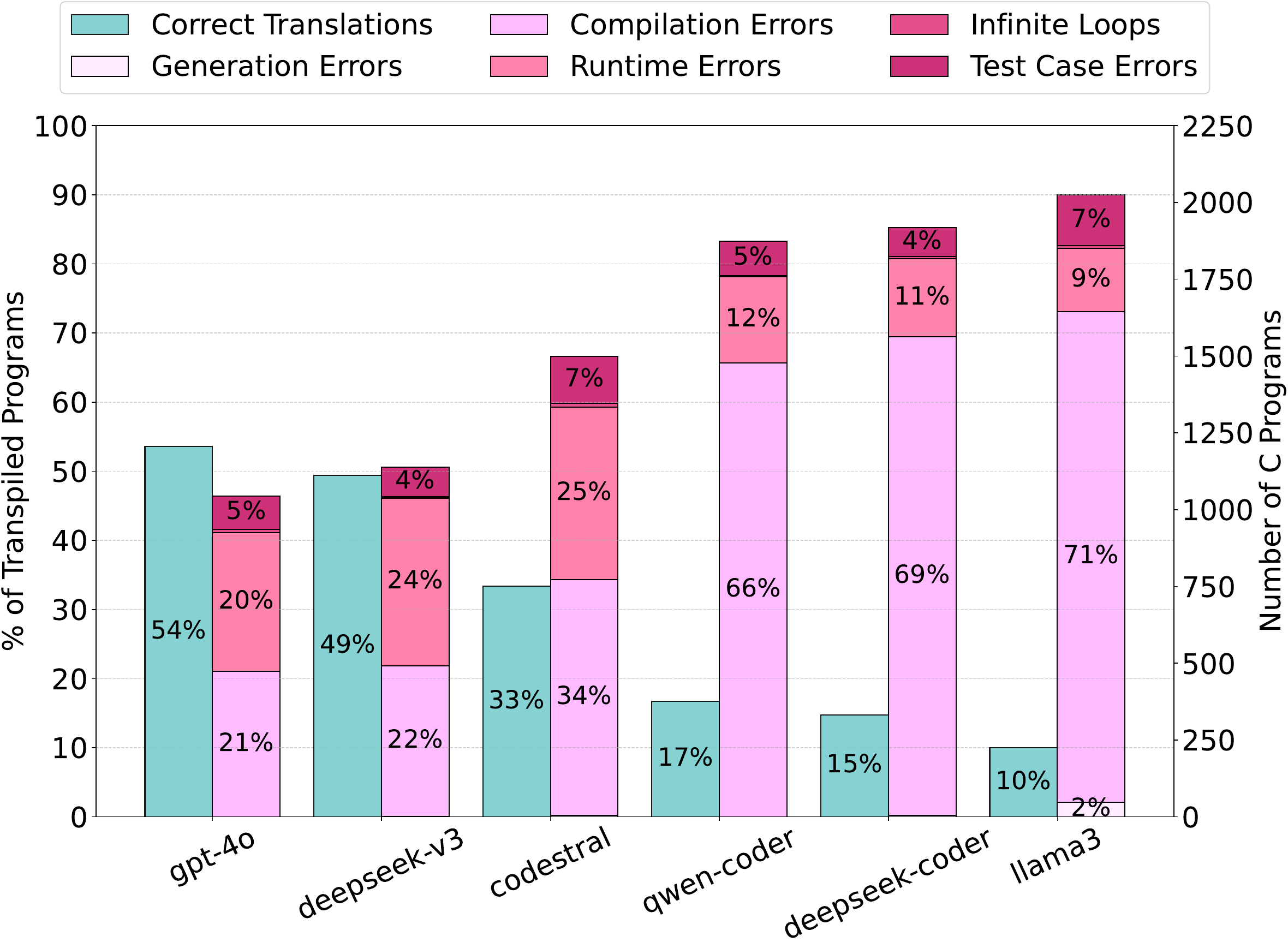}  % Adjust width as needed
    \vspace{-3ex}
    \caption{Percentage of correct Rust translations out of 2,653 C programs,
    and breakdown of the different types of failures for unsuccessful
    translations.}
    \Description{A grouped bar chart comparing six LLMs showing correct translation percentages on the left bars and stacked error breakdowns on the right bars. Error categories include generation errors, compile errors, runtime errors, infinite loops, and test case failures.}
    \label{fig:base_results}
\end{figure}

Figure~\ref{fig:base_results} presents a comparative analysis of the six evaluated
LLMs in their ability to perform basic C-to-Rust translation (without any
attempt to fix any errors). For each
model, the left bar in the group represents CA, while the right stacked bar
illustrates a breakdown of the encountered translation errors. Following the categorization of
Pan et al.~\cite{pan2024lost}, we classify these errors into five distinct
types: 
\ding{172}~\textit{Generation Error}: The model either fails to generate a response according to required format, or the prompt exceeds the LLM's context length.
\ding{173}~\textit{Compilation Error}: The transpiled Rust code fails to compile.
\ding{174}~\textit{Runtime Error}: The Rust code is compiled successfully, but the execution of the program fails (e.g., panics)
\ding{175}~\textit{Infinite Loop}: The program enters a non-terminating loop.
\ding{176}~\textit{Test Case Error}: Execution of at least one test case fails.
% \begin{itemize}
%     \item \textit{Generation Error}: The model either fails to generate a response according to
%     required format, or the prompt exceeds the LLM's context length. 
%     \item \textit{Compilation Error}: The transpiled Rust code fails to compile. 
%     \item \textit{Runtime Error}: The Rust code is compiled successfully, but the
%     execution of the program fails (e.g., panics). 
%     \item \textit{Infinite Loop}: The program enters a non-terminating loop. 
%     \item \textit{Test Case Error}: Execution of at least one test case fails.
% \end{itemize}

Among the evaluated models, \texttt{gpt-4o} and \texttt{DeepSeek-V3} achieve the highest base
computational accuracy (54\% and 49\%, respectively). 
% Although
% \texttt{DeepSeek-V3} is an open-source model, DeepSeek provides API access to its largest variant at
% a cost approximately 25 times lower than \texttt{gpt-4o}, while delivering comparable
% performance. 
% In contrast to larger LLMs, smaller open-source models perform
% significantly worse, with much higher error rates.
As evident from the stacked bars, compilation errors are the most
frequent failure mode across all models. 

% On average, the smaller open-source code models in our study demonstrate over
% twice the CA of \texttt{Llama3}. Despite its 70B parameter size and substantial
% resource demands, \texttt{Llama3}'s general-purpose design results in limited performance on
% code-specific tasks, underscoring the importance of domain-specific training for
% programming language translation. 
% Interestingly, despite having only seven
% billion parameters, \texttt{Qwen2.5-Coder} achieves performance on par with much larger competitors.

% As evident from the stacked bars, compilation errors are the most
% frequent failure mode across all models. 
% This highlights that syntactic correctness
% is a key challenge in LLM-based translation. The particularly high prevalence of
% compilation errors in smaller open-source models further emphasizes their
% limited understanding of Rust's syntax and compilation rules.
% For \texttt{gpt-4o} and \texttt{DeepSeek-V3}, runtime errors are equally frequent as compilation
% errors, which motivated us to also explore iterative repair techniques
% tailored to runtime errors.

\subsection{RQ2: Compilation Error Analysis}
% Explain the collection of compile errors
% filtered out the top errors
% explain the heatmap
% add the results of iterative compilation repairing
% Create two new metrics repair_rate and pass_rate to explain iterative and
% guided repairing.

\subsubsection{Error Distribution}

A high frequency of compilation errors demands an in-depth study of their root
causes. The transpiled Rust programs result in a diverse range of 
errors, but we focus on the most frequent ones that comprise the vast
majority of cases, since they represent the core
features of Rust with which LLMs struggle. Additionally, we only consider the
compilation errors for which \texttt{rustc} provides specific error codes, because this makes
the categorization of errors easier for systematic analysis.

% The heatmap of \figref{fig:error_distro} illustrates the distribution of the
% union of the top-10 most frequent Rust compilation errors per LLM.
% % encountered in the transpiled programs.
% Cells with darker color correspond to a higher relative
% contribution to the total number of compilation errors for a given LLM.
We observe consistent occurrence of some
errors across all LLMs. Specifically, errors \texttt{E0277} (\emph{``trait not implemented''}) and
\texttt{E0308} (\emph{``mismatched types''}) are the most prevalent, accounting for over 18\% of
all errors in most models, and up to 30\% for \texttt{Qwen2.5-Coder} and \texttt{Llama3}, suggesting
that LLMs struggle with type inference and trait bounds.
 %when generating Rust code.
% We provide a detailed description of all compilation error codes in Appendix~\ref{sec:errorcodes}.
Beyond these common errors, some models exhibit distinct error patterns. For example, 18.9\% of \texttt{DeepSeek-V3} translations
fail due to \texttt{E0428} (\emph{``duplicate definition''}), while
\texttt{Codestral} and \texttt{DeepSeek-Coder} struggle disproportionately (19--22\%) with \texttt{E0599}
(\emph{``method not found''}). Similarly, \texttt{Qwen2.5-Coder} and \texttt{Llama3} result in high rates
of \emph{``lifetime issues''} and \emph{``borrow conflicts.''}
%  hinting at weaker handling of Rust's memory-safety constraints. 
% The relatively lower variance in
% errors such as E0061 (\emph{``invalid number of arguments in function call'})
% implies that LLMs have captured a decent model of basic programming language
% constructs.

\subsubsection{Iterative Compilation Repair}
To evaluate the effectiveness of SafeTrans' repair
phase, we introduce two metrics:
i)~\emph{Repair Rate:} the percentage of transpiled programs 
    that initially failed to compile, but then were successfully fixed during the
    repair phase, resulting in a compilable program;
ii)~\emph{Pass Rate:} the percentage of repaired programs that run
    successfully and pass all test cases.

\begin{table}[t]
    % \vspace{-2ex}
    \centering
    % \scriptsize
    \small
    \renewcommand{\arraystretch}{1}
    \setlength{\tabcolsep}{2pt}
    \caption{Effectiveness of iterative repairing of failed Rust compilations.}\vspace{-2ex}
    \label{tab:iterative_compile_repair_results}
    \begin{tabular}{lrrrr}
        \hline
        \textbf{LLM} & \textbf{\makecell[r]{Compilation\\Failures}} &
        \textbf{\makecell[r]{Repaired}} &
        \textbf{\makecell[r]{Repair \\ Rate (\%)}} & \textbf{\makecell[r]{Pass
        \\ Rate (\%)}} \\
        \hline
        Qwen2.5-Coder            & 1743 & 1090 & 62.5 & 23.4 \\
        Llama3                   & 1884 & 1096 & 58.1 & 14.1 \\
        DeepSeek-Coder           & 1836 & 1021 & 55.6 & 33.7 \\
        Codestral                &  906 &  778 & 85.8 & 28.9 \\
        GPT-4o                   &  558 &  522 & 93.5 & 43.4 \\
        DeepSeek-V3              &  579 &  520 & 89.8 & 47.1 \\
        \hline
    \end{tabular}
    \vspace{-3ex}
\end{table}   
Table~\ref{tab:iterative_compile_repair_results} provides a breakdown of the
outcomes of the iterative compilation repair phase in terms of repair rate and
pass rate for each LLM.
The Compilation Failures column corresponds to the number of programs that
failed to compile after the base translation. 
%It shows number of files
%with compilation errors (\#Comp. Failed Trans.) from base translation, number of
%repaired files (\#Rep. Trans.) along with Repair and Pass Rates.
We observe that \texttt{gpt-4o} and
\texttt{DeepSeek-V3} achieve high repair rates of 93.5\% and 89.8\%, respectively,
outperforming the other LLMs. Interestingly, despite being significantly smaller,
\texttt{Codestral} performs comparably to larger LLMs (only 7.7\% and 4\% lower than
\texttt{gpt-4o} and \texttt{DeepSeek-V3} in terms of repair rate). 
% demonstrating its strong ability
% to understand and benefit from compiler feedback.
The repair rate of \texttt{Llama3} (58.1\%) is comparable to other smaller
code-oriented LLMs, such as \texttt{Qwen2.5-Coder} (62.5\%) and \texttt{DeepSeek-Coder} (55.6\%), suggesting
that it can effectively leverage repair prompts to fix faulty translations.
The relatively lower pass rates across all LLMs compared to their repair rates
suggest that while these models can interpret compiler feedback and fix
syntactic issues, they often lose sight of the original program intent and
functional equivalence. 
% As noted by Pan et al.~\cite{pan2024lost}, LLMs may
% introduce new errors while resolving existing ones, requiring multiple
% %iterative
% passes to achieve fully correct and functional translations.

\subsection{RQ3: Effectiveness of Guided Repair}

Even after iterative compilation repair, certain compilation errors remain. 
% (see appendix \ref{unresolved} for errors distribution).
% unresolved. \figref{fig:unfixed_error_distro} shows the distribution of
% error types that persist after the completion of the compilation repair phase.
%It is evident that
Many of
the most frequent errors before repair continue to appear in abundance,
which means that
merely providing compiler feedback to the LLM is insufficient for resolving
them. To better understand why LLMs struggle with these persistent errors,
we selected the following ones (top-8) for further investigation:
\ding{172}~\texttt{E0277}: The type does not implement a required trait.
\ding{173}~\texttt{E0308}: Mismatched types encountered.
\ding{174}~\texttt{E0425}: Use of an undeclared name or identifier.
\ding{175}~\texttt{E0599}: Attempted call on a type that doesn't support it.
\ding{176}~\texttt{E0384}: Cannot assign to an immutable variable.
\ding{177}~\texttt{E0282}: Unable to infer enough type information.
\ding{178}~\texttt{E0502}: Cannot borrow as mutable because it is also borrowed as immutable.
\ding{179}~\texttt{E0499}: Cannot borrow as mutable more than once at a time.

For each of these errors, we examine both their successful and failed
repair cases to identify recurring patterns that lead to the error. Based on
these observations, we develop guided instructions that describe the common
causes and provide example fixes where applicable, which are used in our
subsequent guided repair phase.
For example,
error \texttt{E0384} typically occurs when a new value is assigned to an
immutable variable. Our analysis reveals that such common patterns include
reassigning struct instances and variables introduced through pattern matching.
The instructions in our guided repair are tailored to
these patterns and provide examples to assist the LLM in resolving the error.

To evaluate the effectiveness of guided repair in resolving compilation errors, we define the \emph{resolution rate} (RR) metric, which measures the percentage of targeted errors successfully repaired. Overall, guided repair proves highly effective across most error categories. \texttt{Llama3} achieves the highest average RR, resolving over 60\% of common errors, followed closely by \texttt{DeepSeek-V3} and \texttt{Codestral}. Commonly encountered errors such as \texttt{E0384}, \texttt{E0282}, and \texttt{E0425} exhibit resolution rates above 80\% across multiple models, showing that these issues can be fixed reliably when LLMs receive targeted feedback. Even challenging type inference errors like \texttt{E0277} (“trait not implemented”) and \texttt{E0308} (“mismatched types”) achieve an average resolution rate of approximately 59\% and 55\%, respectively, indicating that guided repair effectively improves model performance even on complex errors.

% Table~\ref{tab:guided_repair_results} summarizes the results on the
% effectiveness of guided repair in resolving the remaining compilation errors.
% %after the initial basic repair phase.
% To accurately evaluate its effectiveness, we introduce the
% \emph{resolution rate} (RR) metric, which measures the percentage of targeted errors
% that are successfully repaired---instead of simply indicating whether a file is
% fixed or not. An RR of zero signifies that the target error was not present in
% the test set. Notably, \texttt{Llama3}, despite being a general-purpose LLM, achieves
% high resolution rates across nearly all targeted errors, suggesting that guided
% repair significantly enhances its ability to acquire the knowledge needed for
% effective correction. Errors such as \texttt{E0384}, \texttt{E0282}, and
% \texttt{E0425} consistently show high resolution rates across multiple models,
% indicating that these error types can be fixed easily when LLMs are provided
% with sufficient context and targeted feedback.

\vspace{-2ex}
\subsection{RQ4: Overall Translation Success Rate}
\label{sec:rq4}

\begin{figure}[t]
    \centering
    \includegraphics[width=0.96\columnwidth]{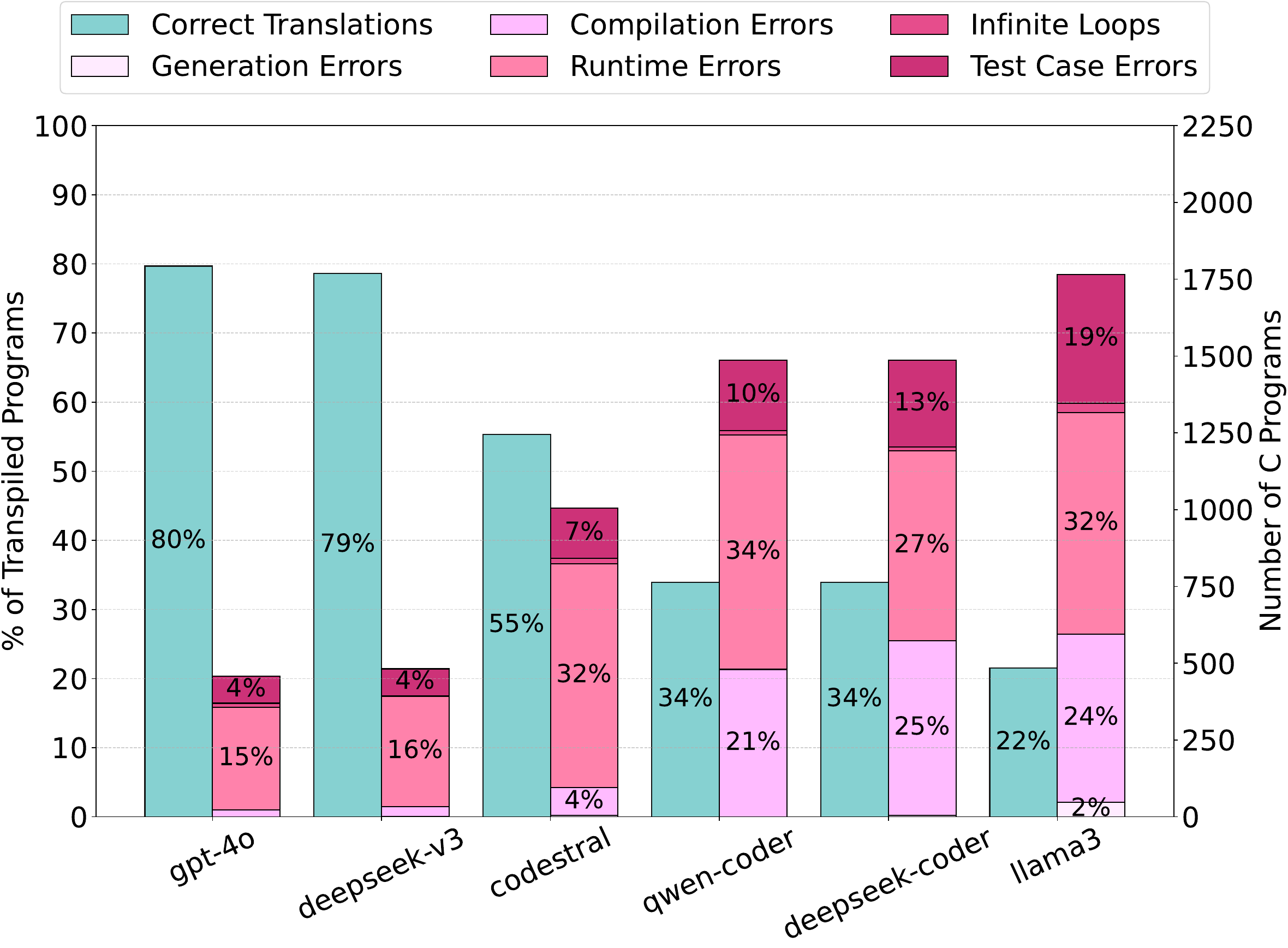}  % Adjust width as needed
    \vspace{-3ex}
    \caption{Final translation success rate and error breakdown for the full SafeTrans pipeline.}
    \Description{A grouped bar chart showing the final translation success rates for six LLMs after the complete SafeTrans pipeline, with stacked bars showing the remaining error types for unsuccessful translations.}
    \label{fig:final_results}
    \vspace{-4ex}
\end{figure}

% \begin{figure*}[t]
%     \centering
%     % \includegraphics[scale=0.5 ]{images/methodology.pdf}  % Adjust width as needed
%     \includegraphics[width=0.90\textwidth]{images/final_results_with_metrics.pdf}  % Adjust width as needed
%     \caption{CDFs of code complexity metrics---lines of code (top) and number
%     of pointers (bottom)---for successful and unsuccessful translation
%     outcomes, as a percentage of the 2,653 C programs in our data set.
%     As the complexity of the code increases, the rate of failed translations
%     also increases.}
%     \label{fig:final_results_with_metrics}
% \end{figure*}

After combining the basic and guided compilation repair approaches, in this
section we evaluate the overall performance of the complete SafeTrans pipeline
in successfully translating C programs into Rust.
Figure~\ref{fig:final_results} illustrates the substantial improvements achieved
through the iterative repair techniques across all LLMs. The computational
accuracy (CA) of both \texttt{gpt-4o} and \texttt{DeepSeek-V3}
is increased by approximately 25\% compared to
their base CA, reaching 80\% and 79\%, respectively.
Even for the underperforming models,
CA is improved by roughly twice (+22\%
for \texttt{Codestral}, +17\% for \texttt{Qwen2.5-Coder}, +19\% for \texttt{DeepSeek-Coder} and +12\% for \texttt{Llama3}).
% Among the smaller open source code-LLMs,
% \texttt{Qwen2.5-Coder} and \texttt{DeepSeek-Coder} achieve the same CA performance (34\%),
% despite \texttt{Qwen2.5-Coder} being smaller (7B) than \texttt{DeepSeek-Coder} (16B).

For \texttt{gpt-4o} and \texttt{DeepSeek-V3}, we observe a drastic reduction across all error types,
particularly in compilation errors, which drop below 2\% (from 22--23\% in
Figure~\ref{fig:base_results}). This indicates that
larger LLMs are highly effective at debugging erroneous programs when provided
with appropriate feedback. Similarly, compilation errors for
\texttt{Codestral}, \texttt{Qwen2.5-Coder}, \texttt{DeepSeek-Coder}, and \texttt{Llama3} drop by
30, 45, 44, and 47 percentage points, respectively.
Among these, \texttt{Llama3} exhibits the largest reduction in
compilation errors, indicating that even general-purpose LLMs can
substantially benefit from our iterative repair strategies.
% to enhance their code understanding capabilities.
For the smaller LLMs, we can see that there is
an increase in non-compilation errors. This occurs because programs that
previously failed to compile,
once repaired, may generate new runtime or logic errors due to
the hallucination tendencies of LLMs. However, the overall increase in CA shows
the potential of targeted iterative repair techniques across models.    

\subsection{RQ5: Scalability on Complex Programs}
\label{sec:rq5}

To assess SafeTrans's scalability, we evaluated it on two real-world libraries,
\texttt{url\_parser}~\cite{url-parser} and \texttt{avl\_tree}~\cite{avl-tree},
which have been used in prior studies~\cite{nitin2025c2saferrust,
li2024userstudy, sactor}. While existing tools such as C2SaferRust, VERT,
FLOURINE, and SACTOR either fail to compile these programs or produce
translations with extensive unsafe code, SafeTrans successfully compiles both
libraries while avoiding unsafe constructs. However, verification remains
partial, as test-based feedback alone is insufficient to pinpoint semantic
errors. A detailed comparison is provided in Appendix~\ref{sec:tool_comparison}
(Table~\ref{tab:comparison}).

%% file: vulnerability.tex
\section{Vulnerability Analysis}
\label{sec:vulnerability}
% \vspace*{0.8ex}

\subsection{Identification of Potential Vulnerabilities}

A major goal of our study is to assess the extent to which any vulnerabilities
present in the original C code are effectively mitigated in the translated
Rust code.
Instead of planting bugs in existing programs or collecting a different
data set of vulnerable programs, we observe that due to the nature of the
CodeNet data set, its programs already contain numerous flaws that would pose
security risks if they were to be used in production. 

Inspired by the FormAI data set~\cite{tihanyi2023formai},
we used the Efficient SMT-based Context-Bounded Model Checker
(ESBMC)~\cite{gadelha2018esbmc} formal verification tool
to analyze the programs in our CodeNet data set and
identify various types of vulnerabilities in them.
% such as illegal memory accesses and integer overflows.
ESBMC uses bounded model checking, which examines the correctness of
a program by converting it into a finite state transition model and exploring
possible states (up to a predefined boundary). 
% ESBMC is an open-source tool that supports multiple programming languages, including C. 
It
automatically verifies both predefined safety properties (e.g., out-of-bounds
array access, illegal pointer dereferences, overflows)
and user-defined program assertions.
We should note that the flaws reported by ESBMC represent
\emph{potential} vulnerabilities---determining whether they are indeed
exploitable is outside the scope of this work. For the sake of brevity, we
refer to these flaws simply as \emph{vulnerabilities} in the rest of this section.
% as the majority of these flaws are indeed critical.
%  (as discussed in Section~\ref{subsec:vul_behavior}).

The outcome of scanning a program with ESBMC can be categorized into one of the
following three cases:
i)~\emph{Verification Successful:} indicates that no
flaws have been found within the defined bounds;
ii)~\emph{Verification Failed:} ESBMC detected one or more
flaws in the target program;
iii)~\emph{Scan Error:} during verification, ESBMC may crash or timeout.
%In such cases, we mark the file as having a Scan Error.
% Table \ref{tab:verification-results} shows the breakdown of ESBMC verification
% results on our evaluation data set. 
Approximately 70\% of
the programs failed verification, i.e., they contain some form of flaw that was not reported by the
authors of CodeNet. 
% We employ the same ESBMC configurations suggested by Tihanyi
% et al.~\cite{tihanyi2023formai}, as our focus is on the behavior of potential
% vulnerabilities that are mitigated as the code is transpiled from C to Rust, rather than reporting all possible
% bugs in a C program. 
As pointed out by
previous works~\cite{tihanyi2023formai, gadelha2018esbmc}, ESBMC cannot produce false positives
or false negatives, as each identified issue is validated by counterexamples, and the fact that successful
verification only occurs up to a predefined bound. This means that the
possibility of some bugs hiding deep in the program still exists, but as we
show, ESBMC still identifies plenty of potential vulnerabilities in the tested programs.
% mikepo: I'm not sure about the following sentence - I don't think we can
% guarantee that no bugs remain in the program, which is what this sentence
% implies:
%
%, and we demonstrate in
%Section~\ref{subsec:vul_behavior} how translating the vulnerable code into
%Rust eliminates this possibility.

% \begin{table}[t]
% \centering
% \caption{Categorization of the outcome of running
% ESBMC~\cite{gadelha2018esbmc} verification on the C programs in our data set.}
% % \vspace{-2ex}
% \label{tab:verification-results}
% \begin{tabular}{lrr}
%     \toprule
%     \textbf{Type} & \textbf{Frequency} & \textbf{Percentage (\%)} \\
%     \midrule
%     Verification Failed & 1906 & 71.8 \\
%     Verification Successful & 332 & 12.5 \\
%     Scan Error & 415 & 15.6 \\
%     \midrule
%     Total & 2653 & 100.0 \\
%     \bottomrule
% \end{tabular}
% \end{table}

\begin{table}[t]
    % \vspace{-1ex} % reduce space above table
    \centering
    \small
    \renewcommand{\arraystretch}{1}
    \setlength{\tabcolsep}{1pt}
    
    \caption{Distribution of the types of detected vulnerabilities in the original C programs, as provided by ESBMC~\cite{gadelha2018esbmc}.}
    \vspace{-2ex}
    \label{tab:vulnerability-distribution}
    \begin{tabular}{lr}
        \toprule
        \textbf{Vulnerability Type} & \textbf{Instances} \\
        \midrule
        Buffer Overflow & 3,258 \\
        Array Bounds Violated & 2,951 \\
        Arithmetic Overflow & 2,859 \\
        Dereference Failure: NULL Pointer & 753 \\
        Division by Zero & 196 \\
        VLA Array Size Overflows Address Space & 175 \\
        Dereference Failure: Forgotten Memory & 100 \\
        Dereference Failure: Invalid Pointer & 47 \\
        % Dereference Failure: Invalidated Dynamic Object & 12 \\
        % Dereference Failure: Invalid Pointer Freed & 5 \\
        % Dereference Failure: Misaligned Access to Data Object & 4 \\
        \bottomrule
    \end{tabular}
    \vspace{-4ex}
\end{table}

Table~\ref{tab:vulnerability-distribution} shows the distribution of the most
frequently reported types of vulnerabilities in the original C programs as
reported by ESBMC. A
single program can contain multiple vulnerabilities across multiple types.
% Since our evaluation data set comprises competitive style
% programs that mostly involve data supplied through \texttt{stdin} or simple
% files, they commonly use simple \texttt{scanf()} calls, arrays, and use of arithmetic
% operations, and we thus observe many vulnerabilities associated to these
% operations.
Overall, ESBMC identified a total of 10,375 vulnerabilities in the
2,653 programs (about five per program).
% on average).

\subsection{Vulnerability Mitigation}
\label{subsec:vul_behavior}
Despite explicit instructions to generate safe Rust code, all the evaluated
models still produced very few translations containing some \texttt{unsafe} Rust code blocks
(0.89\% for \texttt{DeepSeek-Coder}, 1.25\% for \texttt{Llama3}, 2.0\% for \texttt{Qwen2.5-Coder}, 1.8\% for
\texttt{gpt-4o}, 3.4\% for \texttt{Codestral}, and 4.4\% for \texttt{DeepSeek-V3}). 
% | Model                 | Mean Unsafe % |
% | :---------------------| ------------: |
% | deepseek-chat         |    36.90% |
% | deepseek_coderV2_lite |    36.61% |
% | codestral             |    35.76% |
% | llama3_70b_inst       |    34.52% |
% | gpt-4o                |    32.76% |
% | qwen_coder            |    31.39% |
Moreover, in terms of average proportion of \texttt{unsafe} lines across all translated programs, the models exhibit roughly comparable behavior, with a mean unsafe ratio of approximately 34\% overall.

For each identified flaw, ESBMC generates a ``proof'' of vulnerability in the
form of program states that trigger the flaw. However, in some complex cases, these states do not show complete execution traces and therefore cannot be directly applied to the transpiled Rust programs. We employ a combination of automated and manual analysis to determine whether the same vulnerabilities in the original C programs can be triggered in the corresponding Rust translations. First, we used a combination of Rust memory-safety
verification and undefined-behavior detection tools, including
\texttt{Rudra}~\cite{rudra},
\texttt{RAPx}~\cite{rapx}, and
\texttt{miri}~\cite{rust_miri}. None of these tools report any vulnerabilities
in the transpiled Rust programs. Next, we perform manual analysis to identify
any cases missed by automated checkers. Since scaling this manual analysis to
all translations is not possible, we only focus on arithmetic overflow and null pointer dereference  vulnerabilities.

\textbf{Arithmetic Overflows:} Rust provides built-in protections against
arithmetic overflows in debug mode, but does not perform runtime checks in
release mode by default. Although these checks can also be enabled in release
mode, performing an overflow check for every arithmetic expression can
introduce a significant performance overhead in computation-intensive programs.
Rust allows developers to selectively use methods such as
\texttt{checked\_add} and \texttt{checked\_sub} to handle overflows
safely. Only 0.15\% of correct translations across all models use these check
functions. In our manual analysis of transpiled Rust programs corresponding to
C programs with arithmetic overflow vulnerabilities, we found that such
vulnerabilities often propagate to the Rust translations due to the absence of
explicit \texttt{checked} functions.
%%%%%%%%%%%%%%%%%%%%%%%%%%%%%%%%%%%%%%%%%
% ADD EXAMPLE CODE SNIPPETS IF SPACE PERMITS
% also discuss how integer overflows can be leveraged to mount more advanced errors
% add some performance numbers... maybe write a matmul function for integer matrices and time it with and without the integer overflow checks
%%%%%%%%%%%%%%%%%%%%%%%%%%%%%%%%%%%%%%%%%

\textbf{Null Pointer Dereferences:} We observe that the majority of the \texttt{unsafe}
code in Rust translations arises from global variable usage rather than raw pointer
usage (approximately 9.77\% of all translated programs across models contain
at least one raw pointer dereference). However, we still found cases where a
null pointer dereference vulnerability in C propagates to the Rust translation
due to incorrect handling of \texttt{unsafe} blocks.
While translating a C implementation of B+ trees, \texttt{gpt-4o} introduces a bug by incorrectly translating a C for-loop to a Rust range expression which further leads to a segmentation fault.
% \vspace{-1ex}
\begin{lstlisting}[language=C]
// C: Shift elements right when inserting
for (i = leaf->num_keys; i > insertion_point; i--) {...}
\end{lstlisting}
\vspace{-2ex}
\begin{lstlisting}[language=Rust]
// Rust: Empty range when insertion_point < num_keys
for i in ((*leaf).num_keys..insertion_point).rev() {...}
\end{lstlisting}
% \vspace{-1ex}

When inserting at position 0 with \texttt{num\_keys = 1}, the range \texttt{(1..0).rev()} produces no iterations. Consequently, existing elements are not shifted, and \texttt{pointers[1]} remains uninitialized as \texttt{ptr::null\_mut()}. The corruption occurs during range queries, which iterate over all keys and unconditionally dereference their associated pointers.

%% file: limitations.tex
\section{Limitations and Future Work} 

% Our evaluation data set is drawn from CodeNet and real worl C libraries \texttt{url\_parser} and \texttt{avl\_tree}. Since we use the most recent LLMs, it
% is possible that our benchmark programs were included in the training process of these
% models. However, this potential data leakage between our evaluation data set and
% model training does not pose a major threat to the validity of our results. This
% is because these models are not designed specifically for code translation
% tasks, so they may contain independent program samples, but not their functionally
% equivalent code pairs in other languages (and specifically Rust).

% To measure the functional correctness of the transpiled programs, we use the
% commonly accepted approach of test cases, which carries the inherent risk of
% considering a buggy translation as correct. We relied on the test cases provided by our selected datasets.

% CodeNet to assess the functional correctness of translated programs, and
% our measurements show that these test cases achieve high line coverage.
% A limitation of our evaluation is based on relatively simple programs, with
% corresponding test cases mostly in the form of I/O pairs.
% As part of our future work, we plan to evaluate our approach on more complex
% programs that will require more comprehensive test suites---the main challenge
% such a study entails is that it requires a non-trivial experimental setup, and
% the collection of a large enough data set with enough such
% test cases per program.

For our guided repair of compilation errors, we constructed contextual rules
for only the top-eight errors we encountered, and incorporate them in the prompt during the repair
process. As we have shown, this approach yields a high resolution rate for
target errors. It is possible that after addressing the top errors, other
compilation error types may prevent successful translation. Therefore, the set
of custom instructions for guided repair prompts
can be expanded to address additional errors, which would
potentially increase further the success rate of compilation repair.

% It is important to acknowledge that while Rust effectively mitigates many
% traditional memory vulnerabilities found in C, the language presents its own
% unique classes of potential bugs, particularly in the design and implementation
% of safe abstractions around \texttt{unsafe} code \cite{zheng2023closer,
% gulmez2023friend} even in safe rust \cite{hassnain2024counterexamples}. Future
% work might explore automated methods to refactor these \texttt{unsafe}
% implementations in Rust's translations into safe alternatives where possible.

%% file: conclusion.tex
\section{Conclusion}

We presented SafeTrans, a framework that leverages LLMs
to automate the transpilation of C code into Rust.
Our approach combines basic repair and few-shot guided repair to address the
inherent challenges in translating C code to \emph{idiomatic} and \emph{safe} Rust. Our
extensive experimental evaluation results demonstrate that
SafeTrans offers significant improvements in both
computational accuracy (up to 25\%) and compilation error repairing (up to
93.5\%) compared to basic LLM transpilation.
Our analysis of the security implications of the
transpilation process highlights that LLM generated translations require further processing to mitigate source program vulnerabilities.
We believe our findings highlight the potential of LLMs for
automated transpilation to memory-safe languages and will encourage further
research in this area.
% Our analysis of the security implications of the
% transpilation process highlights how Rust's safety guarantees can mitigate the
% memory vulnerabilities present in the original C programs.
% We believe our findings highlight the potential of LLMs for
% automated transpilation to memory-safe languages and will encourage further
% research in this area.

%% file: appendix.tex
\section{Appendix}
\label{sec:appendix}
\subsection{Comparison with Prior C-to-Rust Translation Tools}
\label{sec:tool_comparison}

Table~\ref{tab:comparison} 
summarizes each tool's reported
performance across three metrics: compilation success, complete code safety,
and verification completeness. These include mostly
prototype tools that require non-trivial setup, and in some cases manual intervention
in appropriately formatting the source programs. Therefore, the presented results have been
gathered directly from the respective publications
describing each tool, rather
than from independent experimentation.
% BEGIN: Original two-column table

% END: Original two-column table

\begin{table}[h]
% \centering
\scriptsize
% \arraystretch{1.1}
\setlength{\tabcolsep}{2pt}
\caption{Comparison of C-to-Rust translation tools on the \texttt{url\_parser} (UP) and \texttt{avl\_tree} (AT) programs.}
\label{tab:comparison}
\vspace{-1ex}
\begin{tabular}{p{1.4cm} p{1.1cm} l c c p{2.8cm}}
\toprule
\textbf{Tool} & \textbf{Model} & \textbf{Prog.} & \textbf{Comp.} & \textbf{Safe} & \textbf{Comments} \\
\midrule
C2SaferRust & gpt-4o-mini & UP & $\checkmark$ & $\times$ & Reports unsafe lines; no test coverage mentioned. \\
\midrule
VERT & gpt-4 & UP & $\times$ & -- & Complete translation fails to compile. \\
     & claude-2 & AT & $\checkmark$ & -- & Only subset of functions translated. \\
\midrule
FLOURINE & claude-3.5 & UP & $\times$ & -- & Complete translation fails to compile. \\
\midrule
SACTOR & gpt-4o & UP & $\checkmark$ & $\times$ & Partially idiomatic; fails verification. \\
       & gpt-4o & AT & $\checkmark$ & $\times$ & Partially idiomatic; fails verification. \\
\midrule
\textbf{SafeTrans} & gpt-4o & UP & $\checkmark$ & $\checkmark$ & Passes partial verification. \\
                   & gpt-4o & AT & $\checkmark$ & $\checkmark$ & Passes partial verification. \\
\bottomrule
\end{tabular}
\end{table}